\documentclass[pdflatex,sn-mathphys-num]{sn-jnl}

\usepackage{graphicx}%
\usepackage{multirow}%
\usepackage{amsmath,amssymb,amsfonts}%
\usepackage{amsthm}%
\usepackage{mathrsfs}%
\usepackage[title]{appendix}%
\usepackage{xcolor}%
\usepackage{textcomp}%
\usepackage{manyfoot}%
\usepackage{booktabs}%
\usepackage{algorithm}%
\usepackage{algorithmicx}%
\usepackage{algpseudocode}%
\usepackage{listings}%
\usepackage{caption}
\usepackage{ulem}

\usepackage{geometry}
\theoremstyle{thmstyleone}%

\theoremstyle{thmstyletwo}%

\theoremstyle{thmstylethree}%

\begin{document}

\title[Article Title]{Crowdsourced reviews reveal substantial disparities in public perceptions of parking}

\author*[1]{\fnm{Lingyao} \sur{Li}}\email{lingyaol@umich.edu}
\equalcont{These two authors contributed equally to this work.}

\author*[2]{\fnm{Songhua} \sur{Hu}}\email{hsonghua@mit.edu}
\equalcont{These two authors contributed equally to this work.}

\author[3]{\fnm{Ly} \sur{Dinh}}\email{lydinh@usf.edu}

\author[1]{\fnm{Libby} \sur{Hemphill}}\email{libbyh@umich.edu}

\affil*[1]{\orgdiv{School of Information}, \orgname{University of Michigan}, \orgaddress{\street{105 S State St}, \city{Ann Arbor},  \state{Michigan}, \country{United States}}}
\affil[2]{\orgdiv{Senseable City Laboratory}, \orgname{Massachusetts Institute of Technology}, \orgaddress{\city{Cambridge}, \state{MA}, \country{United States}}}
\affil[3]{\orgdiv{School of Information}, \orgname{University of South Florida}, \city{Tampa}, \state{FL}, \country{United States}}

\abstract{Due to increased reliance on private vehicles and growing travel demand, parking remains a longstanding urban challenge globally. Quantifying parking perceptions is paramount as it enables decision-makers to identify problematic areas and make informed decisions on parking management. This study introduces a cost-effective and widely accessible data source, crowdsourced online reviews, to investigate public perceptions of parking across the U.S. Specifically, we examine 4,987,483 parking-related reviews for 1,129,460 points of interest (POIs) across 911 core-based statistical areas (CBSAs) sourced from Google Maps. We employ the Bidirectional Encoder Representations from Transformers (BERT) model to classify the parking sentiment and conduct regression analyses to explore its relationships with socio-spatial factors. Findings reveal significant variations in parking sentiment across POI types and CBSAs, with \textit{Restaurants} showing the most negative. Regression results further indicate that denser urban areas with higher proportions of African Americans and Hispanics and lower socioeconomic status are more likely to exhibit negative parking sentiment. Interestingly, an opposite relationship between parking supply and sentiment is observed, indicating increasing supply does not necessarily improve parking experiences. Finally, our textual analysis identifies keywords associated with positive or negative sentiments and highlights disparities between urban and rural areas. Overall, this study demonstrates the potential of a novel data source and methodological framework in measuring parking sentiment, offering valuable insights that help identify hyperlocal parking issues and guide targeted parking management strategies.}


\maketitle

\section{Introduction}\label{sec1}

Due to increased reliance on private vehicles and growing travel demand, parking has been a significant urban issue for decades \cite{shoup2021high}. In the U.S., as of 2022, there were approximately 283 million registered vehicles, with a 3.5\% increase over five years \cite{FHWA2022}. In addition, 91.7\% of U.S. households owned at least one vehicle, and 86.7\% of person trips were made by private cars, making it the predominant transportation mode in the country \cite{NHTS2022}. This heavy reliance on private vehicles has placed immense demands on parking, particularly in urban areas with high trip attractions. However, the dense land development in urban areas leave limited space for parking, exacerbating the supply-demand imbalance in big cities. 

A 2017 parking survey by INRIX, involving nearly 6,000 participants, found that Americans spend an average of 17 hours per year searching for parking, costing each driver an estimated \$345 in wasted time, fuel, and emissions. Additionally, 63\% of respondents reported avoiding driving to destinations due to parking challenges, significantly affecting local businesses and economic activity \cite{INRIX2022}. Therefore, parking management is crucial in urban planning and traffic management, as it directly impacts city functionality and livability \cite{litman2016parking, shoup2006cruising}. Proper management strategies can mitigate parking pains, reduce traffic congestion, support local businesses, and foster urban development. Ensuring parking accessibility and affordability also helps promote social equity \cite{litman2020parking, parmar2020study}, and improving the parking environment can help improve urban aesthetics, cleanness, and safety \cite{shaffer1985perceptions}. Conversely, improper parking management can worsen traffic congestion \cite{shoup2006cruising}, waste time and fuel, increase air pollution \cite{javaid2020determinants}, and discourage travelers from frequenting local businesses \cite{mondschein2020using}.

Over the years, urban planners have invested significant effort into parking management through strategies such as increasing parking supply, enhancing parking efficiency, and reducing parking demand. However, a major challenge is effectively measuring public perception of parking. Accurately quantifying these perceptions is essential for efficient parking management, as it allows decision-makers to pinpoint problem areas and devise community-specific solutions \cite{golias2002off, simicevic2013effect}. On the one hand, detailed public feedback is instrumental in identifying the key factors contributing to local parking issues and helps inform whether improvements should focus on supply, regulation, service, pricing, environment, or facilities \cite{stieffenhofer2016assessing}. Moreover, nuanced feedback can reveal minor or specific issues, crucial for fostering sustainable practices and promoting social equity, ensuring that parking policies are inclusive and considerate \cite{mullan2003you}. On the other hand, public perception can serve as a primary metric to evaluate policy effectiveness. Analyzing changes in public sentiment towards parking before and after the implementation of new management strategies can assess policy success and guide necessary adjustments.

Given the importance of understanding public perception, conventional methods like surveys and interviews have been extensively used to gauge public attitudes toward urban issues \cite{ugolini2022understanding, pakoz2022rethinking}. With the structured questions and quantitative nature, these tools allow urban planners to quantify opinions, pinpoint challenges, and evaluate the overall sentiment regarding urban issues \cite{ugolini2022understanding, pakoz2022rethinking}. In the context of parking, surveys are useful for urban planners as they can be customized to address various aspects of parking, such as availability, convenience, safety, and pricing \cite{long2023towards, anastasiadou2009determining}. However, traditional surveys often suffer from limitations in geographic scope and inherent biases, which can lead to a lack of representativeness in broader planning initiatives. Although large-scale surveys can provide more comprehensive insights, they come with trade-offs, including being time-consuming and costly \cite{jones2013quick}.

Meanwhile, social networks and online review platforms have become increasingly important for people to communicate opinions and feelings. These platforms offer virtual channels that enable faster, more general, and less geographically constrained information dissemination \cite{bendimerad2019user}. By analyzing such information, urban planners can identify emerging challenges \cite{li2020leveraging}, foster public engagement \cite{kleinhans2015using}, and understand community needs \cite{saud2020usage, ilieva2018social}. In the context of parking, one typical example of social media content analysis reveals instances of illegal parking, helping authorities to identify problematic areas and respond accordingly to address traffic issues \cite{kim2018social}. Moreover, the geotagged information in many social media posts and online reviews can help pinpoint specific areas of concern or satisfaction associated with different types of business sites \cite{shelton2015social, song2020using, li2024crowdsourcing}.

When coupled with natural language processing (NLP) tools, social media posts and online reviews can provide valuable insights into public experiences, sentiments, and perceptions of parking-related issues, which are crucial for effective parking management \cite{wu2019enabling, li2022has, hu2023modeling}. One typical application utilizes sentiment analysis derived from Yelp reviews to examine the associations between parking sentiment, management practices, and built environment \cite{jiang2021analyzing}. Mining social media posts and online reviews can also reveal intricate patterns in parking behaviors, such as preferences for specific types of parking facilities \cite{anik2020framework}, peak usage times \cite{baird2022mobile}, and common complaints \cite{arhab2022social}. For example, a prior study analyzing Twitter comments finds that positive reviews often highlight free parking, while negative reviews frequently mention issues like limited spaces and sudden increases in parking fees \cite{arhab2022social}. These NLP-based analyses are instrumental in identifying recurring issues and preferences among drivers expressed on social media, providing administrators with insights into specific and persistent citizen needs \cite{yangdinh2024detection}.

Despite significant advancements in integrating online reviews into urban planning studies, several research gaps remain. First, many existing studies are limited to small regions or specific business types \cite{jiang2021analyzing, mondschein2020using}, which may limit their ability to provide a comprehensive view of parking perception across diverse business types and urban settings. Second, while some research discusses the relationship between public perceptions of parking and local socio-economic factors \cite{jiang2021analyzing}, these studies often rely on descriptive analyses without employing robust statistical models that account for confounding effects, spatial autocorrelation, and random effects. To address these gaps, our study analyzes over 4.98 million parking-related Google Maps reviews from more than 1.13 million points of interest (POIs) across the whole U.S. Via a set of NLP techniques such as Bidirectional Encoder Representations from Transformers (BERT) and Lexical salience-valence analysis
(LSVA), we quantify parking perceptions for each POI, correlate them with local socio-spatial factors, and derive insights from the textual content of the reviews. Specifically, our research seeks to answer three key questions:

\begin{itemize}
    \item RQ1: What disparities exist in public perceptions of parking across (1) POI types and (2) CBSAs?
    \item RQ2: How are these disparities associated with local socio-spatial factors?
    \item RQ3: What are the reasons contributing to different parking sentiment?
\end{itemize}

The first question explores how human-generated content can illuminate public attitudes toward parking across different POI types and CBSAs. Insights from this analysis could deepen our understanding of how public sentiments toward parking vary by location and business type. The second question delves into the relationships between parking sentiment and local socio-spatial factors, aiming to uncover the socioeconomic, demographic, and land development roots of disparities in parking perceptions. The third question focuses on identifying word patterns in reviews to discern practical issues that shape public perceptions of parking as positive or negative. Overall, this study explores the potential of using a new data source—online user reviews—to assess public perception of parking and provides valuable insights to assist urban planners and local administrators in making more informed decisions about parking management.

\section{Methods}\label{sec2}

\subsection{Data preparation}

We choose Google Maps reviews to infer residents' attitudes toward parking. Google Maps is a web mapping platform and consumer application developed by Google, allowing users to freely rate places and share their experiences, feelings, and suggestions about business sites such as restaurants, scenic spots, commercial districts, and airports. We opt for Google Maps data for two primary reasons. First, since 2015, it has experienced a significant increase in reviews, surpassing other platforms like Yelp or TripAdvisor \cite{munawir2019visitor}. Second, compared to social media data like Twitter (now X) or Facebook, Google Maps reviews predominantly reflect customer experiences with businesses, making it a reliable source for our crowdsourcing implementation \cite{lee2018assessment}.

To conduct the analysis, we use a dataset collected and shared by UC San Diego \cite{li2022uctopic, yan2023personalized}. This dataset contains review information from Google Maps (ratings, text, images, etc.), business metadata (address, geographical information, descriptions, category information, price, open hours, and miscellaneous information), and links to related businesses up to September 2021 in the U.S. The dataset comprises a total of 666,324,103 reviews covering 4,963,111 POIs.

To filter reviews implying residents' attitudes toward parking, we select three keywords: ``parking," ``park," and ``parked." However, the word ``park" can be used in a comment to imply a place rather than as a verb for parking. Therefore, we first apply the part-of-speech (POS) tagging to each review and only filter in the reviews where ``park" and ``parked" are used as verbs. As a result, we obtain 4,987,483 reviews mentioning parking-related attitudes covering a total of 1,129,460 POIs for the subsequent analysis.

\subsection{Attitude classification}

The extraction of attitudes toward parking from Google Maps reviews is a text classification task in NLP. Initially, we consider applying established sentiment classification tools such as RoBERTa-based sentiment. However, we find that comments such as ``there are a lot of parking spaces" or ``parking price is reasonable" are classified as neutral. One possible reason is that general sentiment analysis does not capture the specific nuances of attitudes toward parking. Consequently, this necessitates the construction of specialized text classification models tailored to this context.

Compiling classification models requires building training and testing datasets. The attitude classification typically consists of three classes: positive, neutral, and negative. However, we notice that some comments contain relevant keywords but do not express any attitudes toward parking. For example, comments like ``I saw a policeman in the parking lot" or ``The room I booked at the hotel only has a view of the parking lot" are considered unrelated in this context. Therefore, we add a class ``unrelated" to the current output classes. Comments indicating easy accessibility, ample parking spaces, reasonable prices, and friendly employees are generally considered positive. In contrast, comments indicating limited or small parking spaces, high prices, or rude employees are generally considered negative.

Two annotators are involved in the data annotation process. To verify the agreement between the annotators, we randomly select 200 comments, which are labeled independently by each annotator, to calculate the inter-coder agreement. The Krippendorff's $alpha$ is 0.88, which is considered as a satisfactory inter-coder agreement level. Following this way, two annotators label a total of randomly selected 2,000 unique comments from the dataset, with 1,600 used for training and 400 for testing. As a result, the training set contains 738 positive, 79 neutral, 694 negative, and 89 unrelated comments. The testing set contains 190 positive, 165 neutral, 26 negative, and 19 unrelated comments. Specific examples are presented in Appendix \ref{secA1}.

The next step is to compile text classification models. In this study, we include two popular sentiment tools: Vader \cite{hutto2014vader} and RoBERTa-based sentiment \cite{barbieri-etal-2020-tweeteval, loureiro-etal-2022-timelms}. Vader is a rule-based sentiment tool, while RoBERTa uses a robustly optimized BERT pre-training approach built on the transformer architecture with an attention mechanism to generate contextualized representations. Additionally, we include three traditional NLP text classification models based on Term Frequency-Inverse Document Frequency (TF-IDF), in conjunction with three machine learning classifiers, including random forest (RF), stochastic gradient descent (SGD), and logistic regression (LR). We further fine tune a BERT model \cite{devlin2018bert}. This process results in six candidate models, including Vader sentiment, RoBERTa sentiment, TF-IDF + RF, TF-IDF + SGD, TF-IDF + LR, and BERT.

The performance of the six candidate models with tuned hyperparameters is assessed using four metrics: Precision, Recall, F1-score, and Accuracy, as presented in Appendix \ref{secA2}. The BERT model achieves the highest testing accuracy, reaching 0.9, an improvement of 0.28 over the RoBERTa sentiment tool. It also achieves a more balanced and higher F1-score across the output classes. This supports our observation that sentiment analysis alone may not sufficiently capture people's attitudes in specific contexts. Therefore, we apply the BERT model to the entire dataset to extract people's attitudes toward parking. Details for the performance measure is presented in Appendix \ref{secA2}.

\subsection{Socio-spatial regression}
To examine how public sentiments about parking relate to socio-spatial characteristics, we develop a regression model. In this model, we treat public sentiment about parking sourced from Google Maps reviews as the dependent variable analyze it against three sets of independent variables, including socioeconomic, demographic, and land development. These independent variables in each set encompass a wide range of factors, such as household income, population density, age distribution, and land use. Detailed descriptions of these variables are provided in Appendix \ref{secA4}. 

We fit a total of seven models: one incorporating all POIs and six others with each tailored to a specific POI type, including \textit{Restaurant}, \textit{Retail Trade}, \textit{Recreation}, \textit{Personal Service}, \textit{Apartment}, and \textit{Hotel}. Given the extensive range of socio-spatial variables potentially influencing parking sentiment, we conduct variable selection to identify the most relevant set of variables. First, we calculate the Variance Inflation Factor (VIF) to assess multicollinearity, excluding variables with VIFs $>$ 5. Then, we use stepwise regression to select the most significant variables based on the smallest Akaike Information Criterion (AIC) \cite{efroymson1960multiple}. 

To address the modifiable area unit problem, which postulates that the spatial units used for aggregating variables can influence modeling outcomes \cite{hu2022examining}, we conduct our analysis at both the CBSA and Census Block Group (CBG) levels. The study area is limited to the contiguous US, and for each CBG or CBSA, a minimum of 10 parking-related reviews is required to ensure the data adequately represents the regional parking sentiment. A sensitivity analysis on this minimum threshold is detailed in Appendix \ref{secA4}. Note that we only consider CBGs within CBSAs (at least 10,000 population) since parking is mainly an urban issue. CBGs outside of CBSAs typically feature low population densities and are insufficient to experience significant parking issues. Consequently, 86,397 CBGs (accounting for 35.67\% of total CBGs) and 911 CBSAs (accounting for 98.49\% of total CBSAs) are included in the analysis. To explore potential sampling biases, we examine differences in socio-spatial factors between CBGs with and without parking reviews (see Appendix \ref{secA4} for more details). A notable observation is that, due to the nature of Google Maps POIs — which typically receive more reviews in business areas due to higher visitation rates — the parking sentiment analyzed in this study is primarily associated with commercial and workplace activities, and does not encompass residential parking situations.

Next, we employ the Generalized Additive Model (GAM) \cite{wood2017generalized} to fit the regression. This semi-parametric model includes a linear predictor along with a series of additive non-parametric smooth functions of covariates, offering greater flexibility than traditional linear regression. This flexibility is crucial for handling various nonlinear effects through adjustable spline functions within a unified framework \cite{hu2024understanding}. In this study, we include both linear and nonlinear effects. Linear effects capture the relationships between parking sentiment and various socio-spatial factors, while nonlinear effects include a random effect term across all CBSAs to address unobserved within-subject heterogeneity and a spatial interaction term to capture spatial autocorrelation. Further details on the model are presented in Appendix \ref{secA6}.

\section{Results}\label{sec3}

\subsection{How do the public perceptions vary across POI types?}

\begin{figure}[htbp]
  \centering
  \includegraphics[width=1\textwidth]{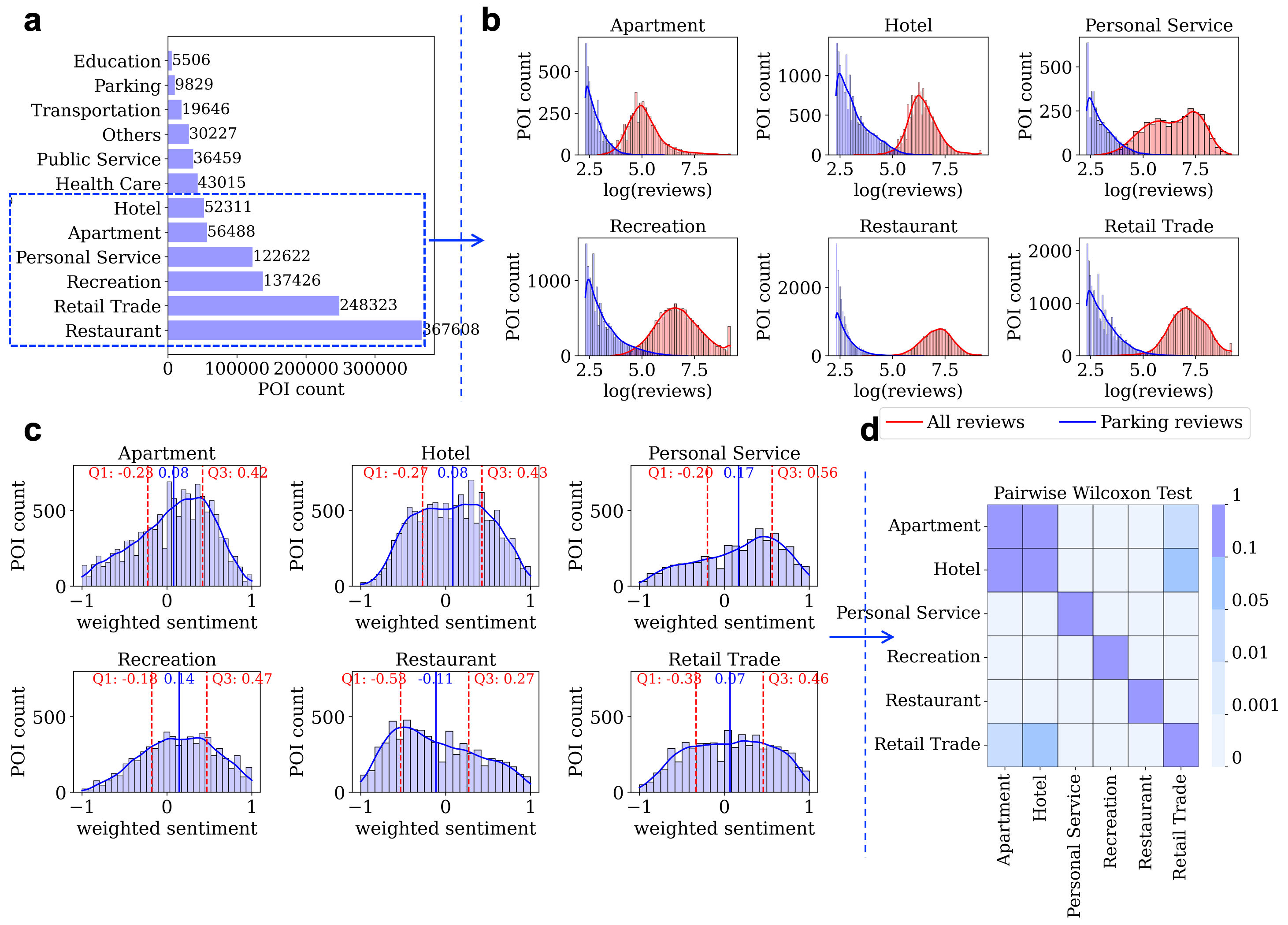}
  \caption{Descriptive results for parking sentiment by POI. (a) Number of POIs by types in the collected dataset (top 10). (b) Distribution of POI count and number of reviews. The x-axis represents the log(reviews), while the y-axis represents the POI count. A threshold of 10 is applied to plot the distribution. (c) Distribution of sentiment. The x-axis represents the weighted sentiment, calculated as the average of all sentiments associated with a POI, while the y-axis represents the POI count. POIs with fewer than 10 parking-related reviews are excluded from the analysis. (d) Pairwise Wilcoxon testing between the sentiment distributions of two POI types.}
  \label{fig:poi_descriptive}
\end{figure}

In response to RQ1(1), we present the descriptive results categorized by POI types in our dataset, as illustrated in Figure \ref{fig:poi_descriptive}. According to Figure \ref{fig:poi_descriptive}(a), the six most frequently mentioned POI types are \textit{Restaurant}, \textit{Retail Trade}, \textit{Recreation}, \textit{Personal Service}, \textit{Apartment}, and \textit{Hotel}. Despite their frequent mentions in reviews, we select these six major POI types for subsequent analysis for two additional reasons. First, they are often tied to daily activities that generate high trip volumes, accounting for over 90\% of total visits across all POIs. Second, business sites such as restaurants, retail trades, and recreation centers typically experience high parking turnover rates and busy demand periods, indicating that parking issues are more pronounced at these POIs. Figure \ref{fig:poi_descriptive}(b) illustrates the distribution of total reviews and parking-related reviews for each POI type. The gap between these two distributions for \textit{Restaurant} POIs is noticeably wider than for other POI types, indicating that only a small proportion of \textit{Restaurant} reviews mention parking. In contrast, the gap between \textit{Apartment} and \textit{Personal Service} is much narrower, suggesting that a higher proportion of reviews for these POI types are related to parking.

Figure \ref{fig:poi_descriptive}(c) displays the sentiment distribution for each POI type. \textit{Restaurant} exhibits a right-skewed distribution with the lowest mean sentiment, indicating more negative reviews, while \textit{Personal Service} and \textit{Recreation} have the most positive sentiment distributions. \textit{Hotel} and \textit{Retail Trade} display a flatter pattern, with their sentiment distribution between -0.5 and 0.5 closely resembling a uniform distribution. To test the similarity between distributions, a pairwise Wilcoxon test is further conducted, as shown in Figure \ref{fig:poi_descriptive}(d). The results indicate no statistically significant difference between the sentiment distributions for apartments and hotels, possibly because both categories involve living spaces, leading to similar attitudes toward parking requirements. Similarly, \textit{Retail Trade} does not show significant differences compared to other categories. \textit{Recreation} and \textit{Personal Service} are particularly positive, while \textit{Restaurant} are notably negative, highlighting distinct sentiment distributions for these POI types.

\subsection{How do the public perceptions vary across CBSAs?}
\begin{figure}[htb]
  \centering
  \includegraphics[width=1\textwidth]{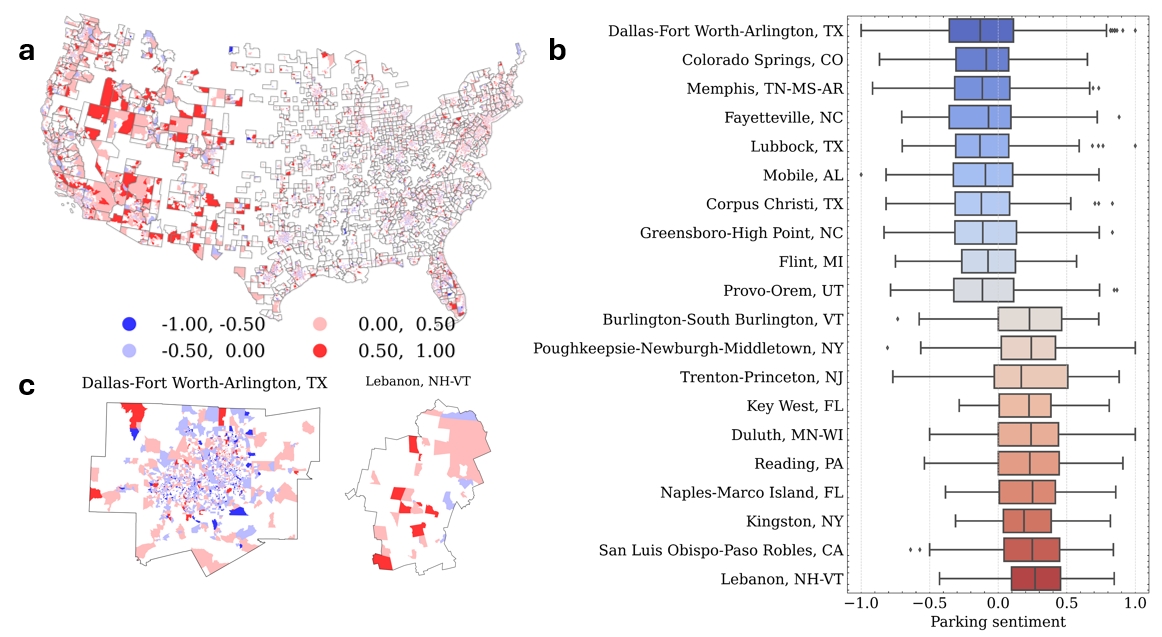}
  \caption{Distribution of parking sentiment across CBSAs in the U.S. (a) Nationwide distribution of parking sentiment, with CBSA boundaries outlined in black. Areas with negative sentiment are shown in blue, while areas with positive sentiment are shown in red. (b) Boxplot of average parking sentiment for the top and bottom 10 CBSAs, including only those with more than 50 CBGs. (c) Close-up views of the two CBSAs with the most positive and negative parking sentiment, using the same color scheme as in panel (a).}
  \label{fig:spatial_dist}
\end{figure}

To answer RQ1(2), we plot the variance in parking sentiment across all CBSAs in the contiguous U.S., as illustrated in Figure \ref{fig:spatial_dist}. Figure \ref{fig:spatial_dist}(a) displays a choropleth map that highlights significant spatial clustering. This is further evidenced by a global Moran's I statistic of 0.173 (P $<$ 0.001), which indicates a positive homogeneity in the spatial distribution of parking sentiment — namely, areas with high sentiment values are clustered near other high-value areas, and similarly for low values. Additional details on the spatial auto-correlation can be found in the Local Indicators of Spatial Association (LISA) cluster analysis in Appendix \ref{secA5}). 

Figure \ref{fig:spatial_dist}(b) displays a ranked bar chart showing the parking sentiment distribution for the top and bottom 10 CBSAs. Notably, the Dallas-Fort Worth-Arlington metropolitan area (the largest CBSA in Texas) exhibits the most negative sentiment nationwide. A further comparison shows that the top 10 CBSAs with the most negative sentiments have a 41.11\% greater average population size than those top 10 CBSAs with the most positive sentiments, indicating that larger CBSAs tend to exhibit more negative attitudes towards parking. To further demonstrate the variance across different CBSAs, Figure \ref{fig:spatial_dist}(c) provides a detailed view of the two CBSAs with the most contrasting sentiments: Dallas-Fort Worth-Arlington, TX (DFWA), and Lebanon, NH-VT (LNV). We observe significant spatial heterogeneity within DFWA, where central urban areas exhibit more pronounced negative sentiments compared to the suburbs. However, in LNV, sentiments are more uniformly distributed across the area.

\begin{figure}[htb]
  \centering
  \includegraphics[width=1\textwidth]{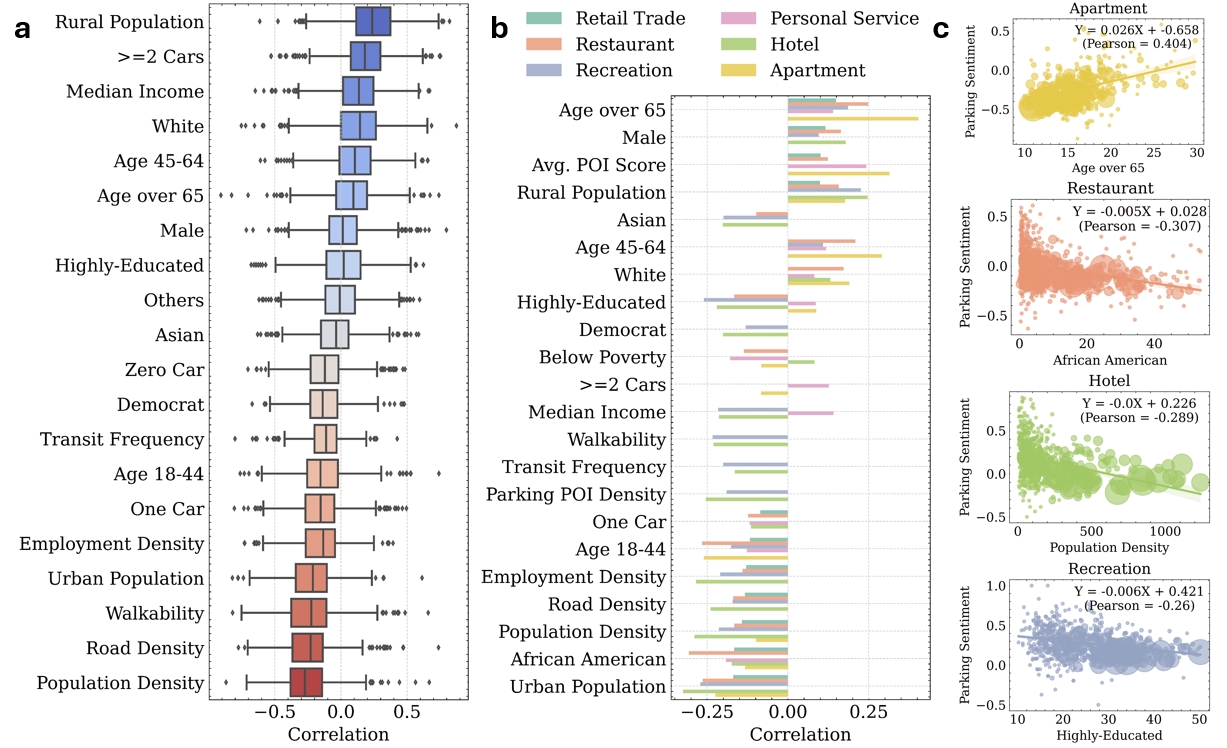}
  \caption{Correlation between parking sentiment and socio-spatial factors, with only correlations having a P-value $<$ 0.001 displayed. (a) Within-CBSA correlation. (b) Between-CBSA correlation segmented by POI types. (c) Scatter plot of univariable regression showcasing the four strongest correlations by POI types, where each point represents a CBSA, sized according to its total population.}
  \label{fig:cbsa_correlation}
\end{figure}

\subsection{How are the perceptions associated with socio-spatial factors?}

In response to RQ2, we first analyze the pairwise correlations between parking sentiment and various socio-spatial factors (Figure \ref{fig:cbsa_correlation}). Two types of correlation are explored: within-CBSA correlation (Figure \ref{fig:cbsa_correlation}(a)), which computes correlations within each CBSA based on its constituent CBGs; and between-CBSA correlation (Figures \ref{fig:cbsa_correlation}(b) and (c)), which aggregates sentiment by CBSA and computes correlations among all CBSAs. Figure \ref{fig:cbsa_correlation}(a) reveals significant rural-urban disparities in parking sentiment. Suburban areas with more rural populations and higher car ownership rates show positive correlations with parking sentiment, while densely populated urban areas with high road density and better walkability show negative correlations.

Figure \ref{fig:cbsa_correlation}(b) extends the analysis by segmenting CBSAs by POI types. As shown, most of the correlations remain robust, but some of them vary considerably across POI types. Factors such as elderly populations, average POI score, and rural population, consistently show positive correlations with parking sentiment across all POI types, while factors such as urban population, African Americans, road density, and population and employment density, consistently display negative correlations. Socioeconomic factors, such as the percentage of highly-educated individuals, the percentage of households below poverty, and median household income, exhibit diverse correlations across different POI types. In Figure \ref{fig:cbsa_correlation}(c), we focus on factors with the strongest correlations and display the scatter plots for several combinations: elderly populations with \textit{Apartment} parking (Pearson = 0.404), African Americans with \textit{Restaurant} parking (Pearson = -0.307), population density with \textit{Hotel} parking (Pearson = -0.289), and education degree with \textit{Recreation} parking (Pearson = -0.260). These scatter plots further underscore that although socio-spatial factors are closely associated with parking sentiment, the relationships are complex and vary significantly across different regions and POI contexts.

However, the correlation analysis is univariable and does not account for confounding effects from other variables. To further investigate RQ2, we fit seven regression models under the GAM framework at the CBG level, with the results presented in Table \ref{tab:gam_outcome}. All coefficients are standardized, making them directly comparable within models. Model goodness-of-fit, the adjusted $R^2$, varies across different POI types, ranging from 0.076 for \textit{Retail Trade} to 0.197 for \textit{Apartment}. The overall adjusted $R^2$ for all POIs is 0.193, which, while moderate, is considered reasonable given the presence of numerous immeasurable factors that may impact public parking sentiment.

Among socioeconomic factors, high population and employment densities exhibit some of the strongest negative correlations with parking sentiment, highlighting the challenges of parking in congested urban centers. Conversely, rural populations present a strong positive relationship with parking sentiment, particularly in the context of \textit{Restaurant} parking, where such a relationship is most pronounced. However, when analyzing \textit{Recreation} parking, a negative relationship is observed between the rural population and parking sentiment, possibly due to the poor parking services at rural recreation spots. Household poverty rates consistently show a negative relationship with parking sentiment, indicating that economically disadvantaged areas face poorer parking conditions. Education levels have an overall negative coefficient, possibly indicating people with higher education attainment have higher expectations or demand for parking. One exception is the \textit{Apartment} parking: Apartments located in highly-educated, low-income areas are more likely to generate positive parking sentiment. This could be because these apartments are often located near universities or are primarily rented by younger households who may not own cars and thus have limited parking needs. This is further supported by the positive relationship between the percentage of zero-car households and the \textit{Apartment} parking sentiment.

Demographic factors show that elderly populations have a consistently positive relationship with parking sentiment, particularly in POIs such as \textit{Apartment} and \textit{Personal Service}. This may reflect elderly adults' less reliance on driving and also parking. Gender-related effects on parking sentiment are nuanced. While no significant relationship emerges when considering all POIs collectively, specific trends appear in certain categories. For instance, \textit{Apartment} shows a positive relationship with gender, while \textit{Hotel} exhibits a negative one. Additionally, racial and ethnic composition is significantly associated with parking sentiment. Compared to areas with predominantly White populations, regions with more Asian populations tend to express more positive parking sentiment. In contrast, areas with more African American and Hispanic populations generally report more negative parking experiences.

Land development factors broadly exhibit complex relationships with parking sentiment. While higher walkability might be expected to correlate with reduced parking demand and more positive sentiment, our results show the opposite. This is possibly due to that the walkability index used in this study is a combination of household entropy, employment entropy, intersection density, and transit proximity (Appendix \ref{tab:variable_summary}). As a result, areas exhibiting higher walkability are more likely to be located in denser urban areas with more negative parking experiences. Similarly, higher transit frequency, which could alleviate the need for parking, doesn't consistently yield significantly positive parking sentiment in our study, since high transit frequency also corresponds to denser urban areas. The density of parking POIs, such as public parking lots and parking garages, also displays a counterintuitive relationship with parking sentiment. While a higher number of parking facilities indicates an increased supply, it also implies greater demand, which can exacerbate parking issues rather than resolve them. Therefore, a negative relationship between parking density and parking sentiment suggests that merely increasing parking supply might not adequately enhance public perception of parking. Lastly, the averaged POI scores provide a direct link to user experiences, with higher POI satisfaction scores also significantly correlating with more positive parking sentiment. 

\begin{table*}[] 
    \centering
    \caption{Results of GAM}
    \label{tab:gam_outcome}
    \resizebox{\textwidth}{!}{%
    \begin{tabular}{lccccccc}
        \hline
        \multicolumn{1}{c}{\textbf{Variables}} & \textbf{All} & \textbf{Restaurant} & \textbf{\begin{tabular}[c]{@{}c@{}}Retail \\ Trade\end{tabular}} & \textbf{Recreation} & \textbf{Hotel} & \textbf{\begin{tabular}[c]{@{}c@{}}Personal \\ Service\end{tabular}} & \textbf{Apartment} \\ \hline
        (Intercept) & \begin{tabular}[c]{@{}c@{}}0.127\\      ***\end{tabular} & \begin{tabular}[c]{@{}c@{}}-0.079\\      ***\end{tabular} & \begin{tabular}[c]{@{}c@{}}0.084\\      ***\end{tabular} & \begin{tabular}[c]{@{}c@{}}0.189\\      ***\end{tabular} & \begin{tabular}[c]{@{}c@{}}0.127\\      ***\end{tabular} & \begin{tabular}[c]{@{}c@{}}0.106\\      ***\end{tabular} & \begin{tabular}[c]{@{}c@{}}-0.348\\      ***\end{tabular} \\ \hline
        \multicolumn{8}{l}{\textbf{Socioeconomics}} \\ \hline
        Population Density & \begin{tabular}[c]{@{}c@{}}-0.033\\      ***\end{tabular} & \begin{tabular}[c]{@{}c@{}}-0.057\\      ***\end{tabular} & \begin{tabular}[c]{@{}c@{}}-0.022\\      ***\end{tabular} & \begin{tabular}[c]{@{}c@{}}-0.056\\      ***\end{tabular} & \begin{tabular}[c]{@{}c@{}}-0.019\\      **\end{tabular} & 0.009 & \begin{tabular}[c]{@{}c@{}}-0.019\\      **\end{tabular} \\ \hline
        Employment Density & \begin{tabular}[c]{@{}c@{}}-0.012\\      ***\end{tabular} & -0.000 & \begin{tabular}[c]{@{}c@{}}-0.023\\      ***\end{tabular} & \begin{tabular}[c]{@{}c@{}}-0.012\\      *\end{tabular} & \begin{tabular}[c]{@{}c@{}}-0.034\\      ***\end{tabular} & \begin{tabular}[c]{@{}c@{}}-0.022\\      **\end{tabular} & 0.011 \\ \hline
        Rural Population & \begin{tabular}[c]{@{}c@{}}0.029\\      ***\end{tabular} & \begin{tabular}[c]{@{}c@{}}0.067\\      ***\end{tabular} & \begin{tabular}[c]{@{}c@{}}0.029\\      ***\end{tabular} & \begin{tabular}[c]{@{}c@{}}-0.015\\      ***\end{tabular} & 0.003 & 0.006 & \begin{tabular}[c]{@{}c@{}}0.053\\      ***\end{tabular} \\ \hline
        Poverty & \begin{tabular}[c]{@{}c@{}}-0.014\\      ***\end{tabular} & \begin{tabular}[c]{@{}c@{}}-0.008\\      **\end{tabular} & \begin{tabular}[c]{@{}c@{}}-0.025\\      ***\end{tabular} & \begin{tabular}[c]{@{}c@{}}-0.014\\      **\end{tabular} & \begin{tabular}[c]{@{}c@{}}-0.011\\      *\end{tabular} & -0.016 & \begin{tabular}[c]{@{}c@{}}0.018\\      **\end{tabular} \\ \hline
        Highly-Educated & \begin{tabular}[c]{@{}c@{}}-0.008\\      ***\end{tabular} & -0.004 & 0.004 & \begin{tabular}[c]{@{}c@{}}-0.048\\      ***\end{tabular} & \begin{tabular}[c]{@{}c@{}}-0.025\\      ***\end{tabular} & 0.010 & \begin{tabular}[c]{@{}c@{}}0.058\\      ***\end{tabular} \\ \hline
        Zero Car & 0.003 & \begin{tabular}[c]{@{}c@{}}0.008\\      **\end{tabular} & 0.007 & 0.001 & \begin{tabular}[c]{@{}c@{}}-0.015\\      **\end{tabular} & -0.016 & \begin{tabular}[c]{@{}c@{}}0.035\\      ***\end{tabular} \\ \hline
        \multicolumn{8}{l}{\textbf{Demographics}} \\ \hline
        Male & 0.000 & 0.005 & -0.004 & -0.004 & \begin{tabular}[c]{@{}c@{}}-0.015\\      ***\end{tabular} & 0.009 & \begin{tabular}[c]{@{}c@{}}0.022\\      ***\end{tabular} \\ \hline
        Age over 65 & \begin{tabular}[c]{@{}c@{}}0.008\\      ***\end{tabular} & \begin{tabular}[c]{@{}c@{}}0.012\\      ***\end{tabular} & \begin{tabular}[c]{@{}c@{}}0.009\\      **\end{tabular} & \begin{tabular}[c]{@{}c@{}}0.012\\      ***\end{tabular} & \begin{tabular}[c]{@{}c@{}}0.011\\      **\end{tabular} & \begin{tabular}[c]{@{}c@{}}0.028\\      ***\end{tabular} & \begin{tabular}[c]{@{}c@{}}0.034\\      ***\end{tabular} \\ \hline
        Age 45-64 & 0.000 & 0.001 & -0.002 & 0.004 & 0.008 & 0.007 & \begin{tabular}[c]{@{}c@{}}0.022\\      ***\end{tabular} \\ \hline
        Asian & \begin{tabular}[c]{@{}c@{}}0.025\\      ***\end{tabular} & \begin{tabular}[c]{@{}c@{}}0.028\\      ***\end{tabular} & \begin{tabular}[c]{@{}c@{}}0.012\\      ***\end{tabular} & \begin{tabular}[c]{@{}c@{}}0.027\\      ***\end{tabular} & \begin{tabular}[c]{@{}c@{}}0.036\\      ***\end{tabular} & \begin{tabular}[c]{@{}c@{}}0.017\\      *\end{tabular} & 0.011 \\ \hline
        African American & \begin{tabular}[c]{@{}c@{}}-0.012\\      ***\end{tabular} & \begin{tabular}[c]{@{}c@{}}-0.016\\      ***\end{tabular} & \begin{tabular}[c]{@{}c@{}}-0.018\\      ***\end{tabular} & -0.002 & \begin{tabular}[c]{@{}c@{}}-0.010\\      *\end{tabular} & 0.000 & \begin{tabular}[c]{@{}c@{}}-0.020\\      **\end{tabular} \\ \hline
        Hispanic & \begin{tabular}[c]{@{}c@{}}-0.012\\      ***\end{tabular} & \begin{tabular}[c]{@{}c@{}}-0.010\\      **\end{tabular} & \begin{tabular}[c]{@{}c@{}}-0.015\\      ***\end{tabular} & -0.010 & -0.005 & -0.018 & \begin{tabular}[c]{@{}c@{}}-0.024\\      **\end{tabular} \\ \hline
        Others & 0.002 & 0.003 & 0.004 & 0.003 & 0.006 & -0.023 & 0.004 \\ \hline
        \multicolumn{8}{l}{\textbf{Land development}} \\ \hline
        Walkability & 0.002 & 0.005 & \begin{tabular}[c]{@{}c@{}}-0.010\\      **\end{tabular} & \begin{tabular}[c]{@{}c@{}}-0.010\\      *\end{tabular} & \begin{tabular}[c]{@{}c@{}}-0.031\\      ***\end{tabular} & \begin{tabular}[c]{@{}c@{}}0.028\\      ***\end{tabular} & \begin{tabular}[c]{@{}c@{}}0.028\\      ***\end{tabular} \\ \hline
        Transit Frequency & \begin{tabular}[c]{@{}c@{}}0.007\\      *\end{tabular} & 0.005 & \begin{tabular}[c]{@{}c@{}}0.016\\      **\end{tabular} & 0.003 & 0.004 & -0.012 & 0.004 \\ \hline
        Parking POI Density & \begin{tabular}[c]{@{}c@{}}-0.011\\      ***\end{tabular} & \begin{tabular}[c]{@{}c@{}}-0.010\\      **\end{tabular} & \begin{tabular}[c]{@{}c@{}}-0.011\\      *\end{tabular} & -0.004 & \begin{tabular}[c]{@{}c@{}}-0.019\\      ***\end{tabular} & -0.007 & -0.014 \\ \hline
        Avg. POI Score & \begin{tabular}[c]{@{}c@{}}0.041\\      ***\end{tabular} & \begin{tabular}[c]{@{}c@{}}0.025\\      ***\end{tabular} & \begin{tabular}[c]{@{}c@{}}0.024\\      ***\end{tabular} & \begin{tabular}[c]{@{}c@{}}0.047\\      ***\end{tabular} & \begin{tabular}[c]{@{}c@{}}0.061\\      ***\end{tabular} & \begin{tabular}[c]{@{}c@{}}0.039\\      ***\end{tabular} & \begin{tabular}[c]{@{}c@{}}0.033\\      ***\end{tabular} \\ \hline
        \multicolumn{8}{l}{\textbf{Nonlinear terms}} \\ \hline
        ti(Lat,Lng) & \multicolumn{1}{l}{8.939} & \multicolumn{1}{l}{7.902} & \multicolumn{1}{l}{10.728} & \multicolumn{1}{l}{3.886} & \multicolumn{1}{l}{2.841} & \multicolumn{1}{l}{2.848} & \multicolumn{1}{l}{6.923} \\ \hline
        s(CBSA) & \multicolumn{1}{l}{340.108} & \multicolumn{1}{l}{203.176} & \multicolumn{1}{l}{126.417} & \multicolumn{1}{l}{112.860} & \multicolumn{1}{l}{174.253} & \multicolumn{1}{l}{43.249} & \multicolumn{1}{l}{63.003} \\ \hline
        \multicolumn{8}{l}{\textbf{Model fit}} \\ \hline
        $R^2$ (Adjusted) & \multicolumn{1}{l}{0.193} & \multicolumn{1}{l}{0.121} & \multicolumn{1}{l}{0.076} & \multicolumn{1}{l}{0.079} & \multicolumn{1}{l}{0.144} & \multicolumn{1}{l}{0.083} & \multicolumn{1}{l}{0.197} \\ \hline
        Sample size & \multicolumn{1}{l}{86397} & \multicolumn{1}{l}{26963} & \multicolumn{1}{l}{21559} & \multicolumn{1}{l}{12106} & \multicolumn{1}{l}{10008} & \multicolumn{1}{l}{4537} & \multicolumn{1}{l}{4602} \\ \hline
        \multicolumn{8}{l}{\begin{tabular}[c]{@{}l@{}}Notes: Significance codes: 0 ‘***’ 0.001 ‘**’ 0.01 ‘*’ 0.05 ‘’ 1. \\In the modeling of all POIs, POI type is controlled as a categorical factor. \\ s() means the spline function. ti() means the marginal nonlinear interaction function.\end{tabular}}
    \end{tabular}%
    }
\end{table*}

\subsection{What are the reasons contributing to parking sentiment?}

\begin{figure}[htbp]
  \centering
  \includegraphics[width=1\textwidth]{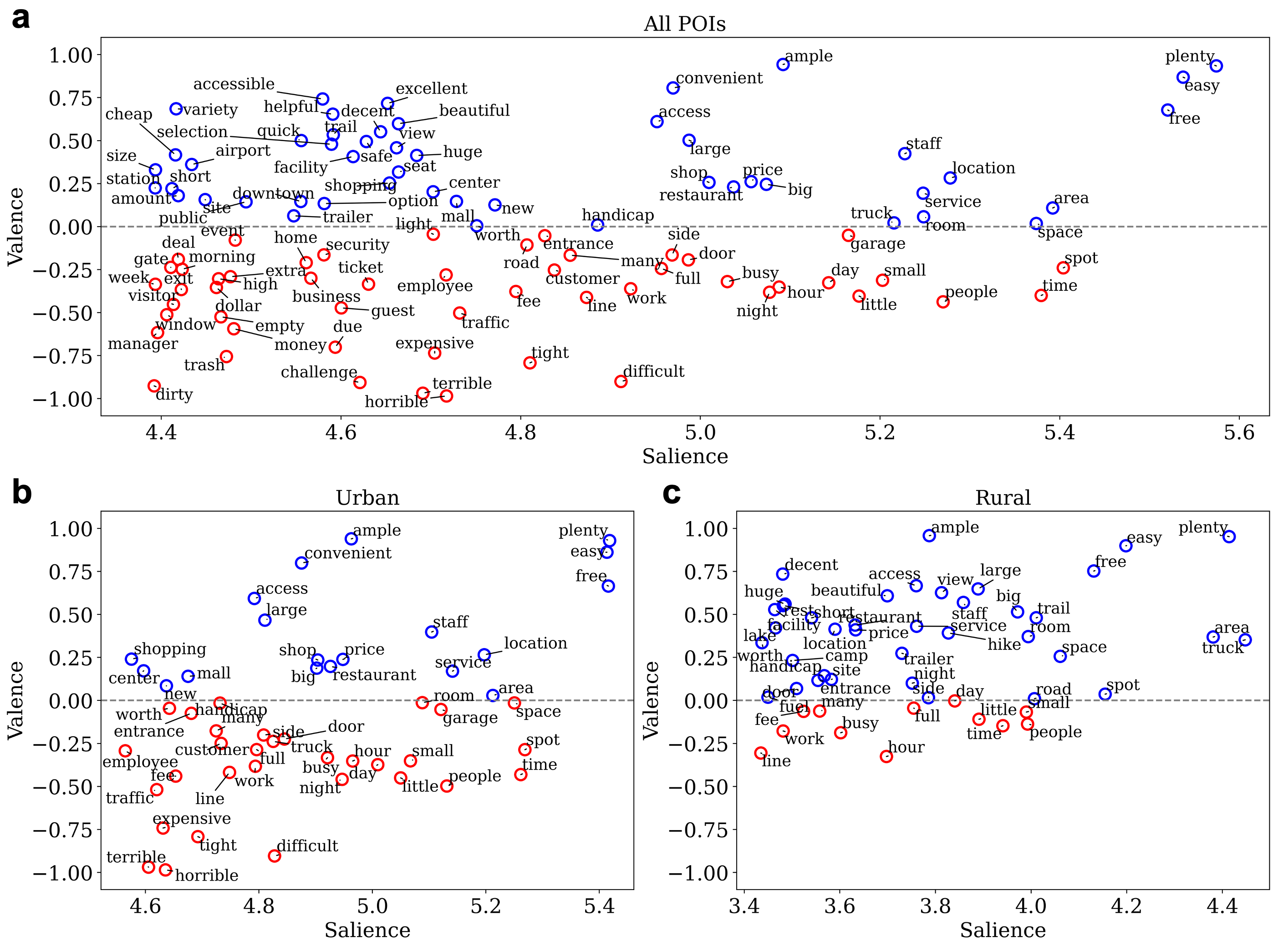}
  \caption{Textual analysis using LSVA based on (a) all POIs, (b) POIs located in the urban areas, and (c) POIs located in the rural areas. Terms on the extremes of the x-axis indicate high or low salience in the reviews, while terms on the extremes of the y-axis indicate strong positive or negative sentiment.}
  \label{fig:textual}
\end{figure}

In response to RQ3, which asks about the textual elements of the reviews that indicate their sentiment on the parking situation, we use a textual analysis technique called Lexical salience-valence analysis (LSVA) \cite{taecharungroj2019analysing}. This tool allows us to show how terms are associated with positive and negative sentiments, as illustrated in Figure \ref{fig:textual}. For all POIs (see Figure \ref{fig:textual}a), positive terms such as ``plenty,'' ``easy,'' ``free,'' ``convenient,'' and ``ample'' appear both frequently and with positive sentiment. These terms suggest that reviewers appreciate the availability and ease of parking when it is abundant and free in cost. Conversely, negative terms like ``small,'' ``little,'' ``difficult,'' ``tight,'' ``dirty,'' ``expensive'' are prominent in use, indicating patrons' concerns for parking conditions relating to limited spaces, high cost, and cleanliness. 

We further examine the differences in terms used in urban (see Figure \ref{fig:textual}b) and rural (see Figure \ref{fig:textual}c) POIs, with the assumption that urban and rural areas face distinct challenges relating to space in general, and parking in particular based on our observation in Section 3.2. Urban areas often have higher population density, more complex infrastructure, and greater competition for limited parking spaces. These factors often lead to higher costs, more frequent traffic congestion, and challenges in finding available parking. Indeed, we observe higher counts of negative terms used in urban POIs compared to rural POIs, where terms such as ``time,'' ``little,'' ``small,'' ``difficult'' are frequently used. Notably, terms like ``time,'' ``little,'' and ``small,'' are also present in the rural POIs' reviews, though with lower salience. That is, patrons in rural areas share the same concerns for parking conditions in similar ways to patrons in urban areas. These concerns include the availability of parking spaces, the convenience of parking locations, and the time required to find parking.


We also find similar terms in both urban and rural POIs that signal positive sentiment regarding parking, namely ``plenty,'' ``easy,'' ``free,'' and ``ample''. However, there are notable differences in terms of the context of the POIs that are discussed in urban and rural areas. In particular, in urban areas, there are a number of terms that are related to service quality in two POI types, ``shopping'' and ``restaurant''. These terms include ``staff,'' ``service,'' and ``location,'' suggesting that patrons in urban areas associate the parking experience with quality of service received. On the other hand, in rural POIs, terms were frequently related to outdoor activities, like ``trail,'' ``hike,'' ``camp,'' and ``site.'' There are also terms relating to trucker activities such as ``truck,'' ``road,'' and ``trailer.''

With the findings above, we have a reason to suspect that there are differences in the salience and valence of parking reviews depending on the POI type. Figure \ref{fig:poi_textual} shows the differences in terms used across six different POI types. Our first observation is that the number of negative terms used in reviews of \textit{Apartment} parking conditions is notably higher compared to other POI types. Additionally, there is a significantly lower number of positive terms used in \textit{Apartment} parking reviews. When positive terms are used, parking is associated with the convenience of the parking spot (e.g., ``easy,'' ``location'') and the amount of parking available. (e.g., ``plenty,'' ``storage''). On the other hand, there is a relatively equitable distribution of positive and negative terms used to review parking conditions at the remaining five POI types. 
The differences in negative terms highlight the varying challenges faced in different POI settings. For instance, in \textit{Restaurant}, \textit{Retail Trade}, and \textit{Recreation} areas, terms like ``traffic'' and ``busy'' are more salient, reflecting common issues with congestion. In contrast, \textit{Apartment} and \textit{Hotel} frequently see negative terms such as ``expensive,'' and ``tight,'' which relate to concerns about high costs and limited space. \textit{Personal Service }locations often mention ``difficult'' and ``full,'' indicating that limited parking space and high demand are common issues in these areas.

\section{Discussion}\label{sec4}

\subsection{Key findings}


This study investigates the U.S. nationwide public perceptions of parking through a novel crowdsourced dataset from Google Maps reviews. The sentiment variances across different POI types, CBSAs, and communities with different socio-spatial factors are examined. Textual analysis is further conducted to directly derive insights from review content. The key findings are listed as follows:

First, public perceptions of parking vary significantly across different POI types. Notably, the overall \textit{Restaurant} POIs tend to receive more negative sentiments compared to other POIs, while \textit{Personal Service} and \textit{Recreation} POIs are associated with more positive sentiments. This variation reflects the differing demands and needs for parking, as illustrated by Figure \ref{fig:poi_textual}. For example, for \textit{Apartment}, public comments often focus on service, price, parking garage availability, and management. For \textit{Restaurant}, the primary concerns are space and time. \textit{Hotel} parking frequently raises issues related to parking fees and night parking.

Second, public perceptions of parking exhibit a strong positive spatial homogeneity but differ remarkably across CBSAs. Larger CBSAs with greater population sizes often have more negative attitudes toward parking: The top 10 CBSAs with the most negative sentiments have a 41.11\% greater average population size than the top 10 CBSAs with the most positive sentiments. Our findings also reveal notable rural-urban disparities in parking sentiment. Suburban areas, which typically have higher car ownership rates and more rural populations, exhibit more positive correlations with parking sentiment. Conversely, densely populated urban areas with higher road density and better walkability generally have more negative parking sentiment. 

Third, strong relationships are documented between public perceptions of parking and different socioeconomic, demographic, and land development factors. Some relationships are consistent. Areas with higher population and employment densities, higher poverty rates, and more African Americans and Hispanics exhibit more negative parking sentiment , while areas with more elderly populations and more Asians show more positive sentiment. These findings document that disparities in parking sentiment may stem from deep-rooted socioeconomic inequities, racial segregation, and urban-rural divides in the U.S. Some relationships, however, vary by POI types, underscoring the importance of developing parking management strategies that are tailored to specific business types. Last, land development factors such as walkability and transit frequency do not always show positive relationships with parking sentiment. Similarly, increasing parking facility density does not necessarily improve sentiment, indicating that simply increasing parking supply is insufficient to address underlying parking issues.

Last, consistent with findings from the prior research questions, we find notable impacts of various factors that distinguish between urban and rural areas on public sentiment regarding parking. We find that in urban areas, where there's higher population density and thus higher competition for spaces such as parking, sentiment is overall more negative. These areas often face higher costs, more frequent traffic congestion, and challenges in finding available parking. The prominence of negative terms such as ``small’’, ``little’’, ``expensive’’ in reviews of urban POIs highlight these challenges. This aligns with existing literature that highlights the greater parking difficulties in urban environments \cite{biswas2017effects}. Additionally, urban reviews often include terms related to service quality in ``shopping’’ and ``restaurant’’ POIs, such as ``staff’’, ``service’’, and ``location’’, indicating that customers in urban areas link their parking experiences with the quality of service they receive \cite{de2013shopping}. In fact, De Nisco and Warnaby (2013) \cite{de2013shopping} find that elements such as the availability of parking, ample spaces for pedestrian walking, and attractive store exteriors are positively correlated with customer perceptions of the overall service quality provided in urban POIs, particularly in shopping areas.

\subsection{Implications}

Our key findings highlight the necessity for designing customized parking solutions to address urban-rural parking disparities. In densely populated urban centers, where negative sentiment towards parking is prevalent but parking spaces are limited, strategies should focus on decreasing parking demand or increasing parking efficiency. Options could include promoting alternative transportation modes like transit and bikesharing, implementing dynamic pricing models to adjust parking rates based on demand, investing in advanced parking technologies such as real-time parking information apps or intelligent parking systems, and encouraging park sharing among businesses during off-peak times \cite{riggs2014dealing, chiara2020policy}. In suburban areas where parking supply is typically sufficient, efforts should focus on enhancing parking services by expanding spaces, improving signage, and ensuring better access to existing facilities. In addition, given the mixed relationships between parking sentiment and land use, policies should balance the local need for parking specified by land uses, ensuring that parking solutions are integrated seamlessly into the broader urban design and functionality.

The significant socio-spatial disparities in parking sentiment highlight the importance of addressing social inequity when designing parking policies. Policymakers should consider a variety of strategies that not only alleviate parking stress but also promote fairness and accessibility for all sociodemographics. In areas with higher poverty rates or significant racial and ethnic diversity, progressive pricing models that adjust parking fees based on income levels could be implemented. Additionally, enhancing lighting and security in public parking areas, especially in lower-income neighborhoods, can boost local vitality by making these areas more safety and attractive. For regions with a substantial elderly population, local governments should ensure that parking facilities are accessible and senior-friendly, including reserving adequate spaces for vulnerable groups. More importantly, engaging local communities in the planning process is essential to gather feedback and understand diverse needs, especially minorities and vulnerable groups, which helps ensure that parking policies are responsive and equitable.

The variance in POI types documented in this study highlights the importance of tailoring parking policies to specific business types. For businesses like restaurants, ensuring parking availability during peak times is essential. Restaurants should prioritize offering sufficient and convenient parking spots to enhance customer satisfaction.  This is particularly important in urban POIs where competition for parking is high, and customer satisfaction is closely tied to service quality aspects, with parking being a major component \cite{de2013shopping, biswas2017effects}. Hotels, on the other hand, need strategies to address parking fee concerns and ensure availability, especially for overnight guests. Additionally, we suggest that commercial property owners incorporate adequate parking facilities into their building plans, particularly in high-demand areas. If parking spaces are limited, collaborating with neighboring businesses to offer shared parking solutions could help mitigate the fluctuating parking demands between idle and busy times \cite{abbott2017utilizing}. Finally, for apartments, we suggest that parking facilities should not only adequate but also well-maintained, addressing critical factors such as service, pricing, and convenience \cite{de2023apartment}. Furthermore, enhancing the management of parking garages and maintaining clear communication regarding parking policies are essential steps toward improving overall parking experiences.

\subsection{Limitations and future work}

This study opens several avenues for future research. First, many POIs have a limited number of parking-related reviews. To ensure robust descriptive analysis based on POI types, we exclude POIs with fewer than 10 parking-related reviews for analysis to avoid small sample size issues. However, this exclusion may overlook valuable insights into parking situations in areas or POI types with sparse reviews. Future research could integrate data from alternative platforms like Yelp, as utilized in previous studies \cite{jiang2021analyzing}, to enrich our dataset.

Although the trained BERT model achieves an overall testing accuracy of 0.9, it still faces challenges in accurately identifying attitudes in reviews, particularly for contextually ambiguous comments. Additionally, our training and testing samples are unbalanced, with fewer samples identified as neutral or unrelated. Consequently, the trained BERT model may have limited ability to effectively interpret neutral or unrelated comments in the context. Future research could consider other large language models (LLMs), such as GPT-4 or LLaMa 3, for more precise text analysis. However, these LLMs come with trade-offs, including higher demands on computing resources, increased costs, and longer processing times.

Next, although leveraging crowdsourced data from online platforms can help reduce biases in selecting participants for an open-ended survey, it still introduces biases based on the individuals who choose to post public reviews. Previous research has shown that younger and more educated individuals are more likely to post online reviews due to their familiarity with social media and online platforms \cite{mellon2017twitter, wang2019demographic}. Additionally, people with predominantly positive or negative experiences are more inclined to leave reviews, which can result in significant variations in the feedback received \cite{filieri2016makes}. More importantly, biases also exist in the locations we analyzed. Since Google Map POIs only include business sites, most of the residential and workplace parking is not represented in this dataset. Hence, additional surveys are still required to fully understand parking sentiments beyond those associated with business locations.

There are two additional avenues that are worth further consideration. First, we could continue collecting data from recent Google Maps reviews to facilitate longitudinal analysis. This insight could allow us to track the longitudinal changes over time, revealing whether parking issues in certain areas with negative sentiments have improved or if areas with positive sentiments have deteriorated. Second, we could extend this framework to investigate public perceptions of other urban issues, such as built environment and health resource accessibility. This expansion could lead to the development of an online platform that integrates various social media and online review sources, which could allow urban planners to comprehensively analyze public perceptions of diverse urban planning challenges.





\section*{Declarations}


\begin{itemize}
\item Conflict of interest/Competing interests. The author declares no competing interests
\item Data availability. The original Google Maps reviews can be downloaded from the website ``Google local review data"" published by researchers from UCSD \cite{li2022uctopic, yan2023personalized}. The processed data can be accessed upon request from the corresponding authors of this paper.
\item Code availability. The code to process and analyze the data is available at \url{https://github.com/Lingyao1219/parking}. 
\end{itemize}

\begin{appendices}

\section{Examples of attitude classification}
\label{secA1}

Table \ref{tab:example} below presents representative examples of attitude classification for parking based on Google Maps reviews. We select the sentence as the unit for attitude classification as some reviews can be long and include aspects not indicative of parking attitude. This fine-grained classification can be helpful when the review conveys opposite sentiments regarding different aspects of a review (see the third example in Table \ref{tab:example}).

\begin{table}[htbp]
  \caption{Representative examples of parking attitude classification based on Google Maps reviews}
  \centering
  \begin{tabular}{p{0.6\linewidth}p{0.2\linewidth}p{0.1\linewidth}}
    \hline
    \textbf{Google Maps reviews} & \textbf{Parking-related reviews} & \textbf{Attitude} \\
    \hline
    Great price. Easy parking. Handicap accessible. & Easy parking. & Positive \\
    \hline
    Great place to host events. There's an attached hotel too if you ever need one. I reckon this would be best for wedding receptions. There's ample parking too! & There's ample parking too! & Positive \\
    \hline
    Parking is poor and expensive. Go early and park off site in a paid lot. The sound at the forum is great! Always a fun, loud show. & Parking is poor and expensive. Go early and park off site in a paid lot. & Negative \\
    \hline
    It's a good art store but they didn't have exactly everything we were looking for such as a variety of wax seal wax, and stamps to go along. They had a great variety of paints and charcoal types though. The location is great but the parking is not so great. A little walk is fine though since it is in little Italy after all. & The location is great but the parking is not so great. & Negative \\
    \hline
    Very gracious and helpful staff! Everything I ordered was delicious! The parking is tight, but it's free. & The parking is tight, but it's free. & Neutral \\
    \hline
    WORST HOTEL!! My check in was delayed 3 hours. The rooms felt damp and had an odd smell to them. The service was not good. The restaurant is not worth it for the price you pay! I was suppose to get a garden view but I got something known as the parking lot view! I also had no hot water to shower with for the first night. Hotel need a big renovation. & I was suppose to get a garden view but I got something known as the parking lot view! & Unrelated \\
    \hline
  \end{tabular}
  \label{tab:example}
\end{table}

\section{Classification performance}
\label{secA2}

For each of the text classification models, we perform hyperparameter tuning within the training partition using K-fold cross-validation with $K=10$. Table \ref{tab:grid_search} shows the parameter of the grid search for each classification model. In each iteration of the training process, the model is trained using nine subsets and validated against the remaining subset. Through 10-fold cross-validation, we test different hyperparameters for each candidate model using grid search and choose the best one for each model.

\begin{table}[htbp]
  \centering
  \caption{Grid search range of candidate classification models}
  \begin{tabular}{p{0.25\linewidth}p{0.25\linewidth}p{0.4\linewidth}}
    \hline
    \textbf{Model} & \textbf{Hyperparameter} & \textbf{Grid search range} \\
    \hline
    Random Forest (RF) &
    \begin{tabular}[t]{@{}l@{}}
      n\_estimators \\
      max\_depth \\
      min\_samples\_leaf
    \end{tabular} &
    \begin{tabular}[t]{@{}l@{}}
      100, 200, 300, 400 \\
      10, 20, 40, 80, 100, 120 \\
      1, 2, 4
    \end{tabular} \\
    \hline
    Stochastic Gradient Descent (SGD) &
    \begin{tabular}[t]{@{}l@{}}
      clf\_\_alpha \\
      clf\_\_max\_iter \\
      clf\_\_penalty
    \end{tabular} &
    \begin{tabular}[t]{@{}l@{}}
      1e-4, 1e-3, 1e-2, 1e-1, 1, 10 \\
      500, 800, 1000, 2000, 3000 \\
      l2, l1, elasticnet
    \end{tabular} \\
    \hline
    Logistic Regression (LR) &
    \begin{tabular}[t]{@{}l@{}}
      clf\_\_C \\
      clf\_\_max\_iter \\
      clf\_\_solver
    \end{tabular} &
    \begin{tabular}[t]{@{}l@{}}
      0.1, 0.5, 1, 2, 5, 10, 20 \\
      10, 20, 50, 100, 200 \\
      sag, saga, lbfgs, newton-cg
    \end{tabular} \\
    \hline
    BERT &
    \begin{tabular}[t]{@{}l@{}}
      num\_\_epochs \\
      batch\_\_size \\
    \end{tabular} &
    \begin{tabular}[t]{@{}l@{}}
      1, 2, 3, 4, 5 \\
      16, 32, 64 \\
    \end{tabular} \\
    \hline
  \end{tabular}
  \label{tab:grid_search}
\end{table}

\begin{figure}[htbp]
  \centering
  \includegraphics[width=1\textwidth]{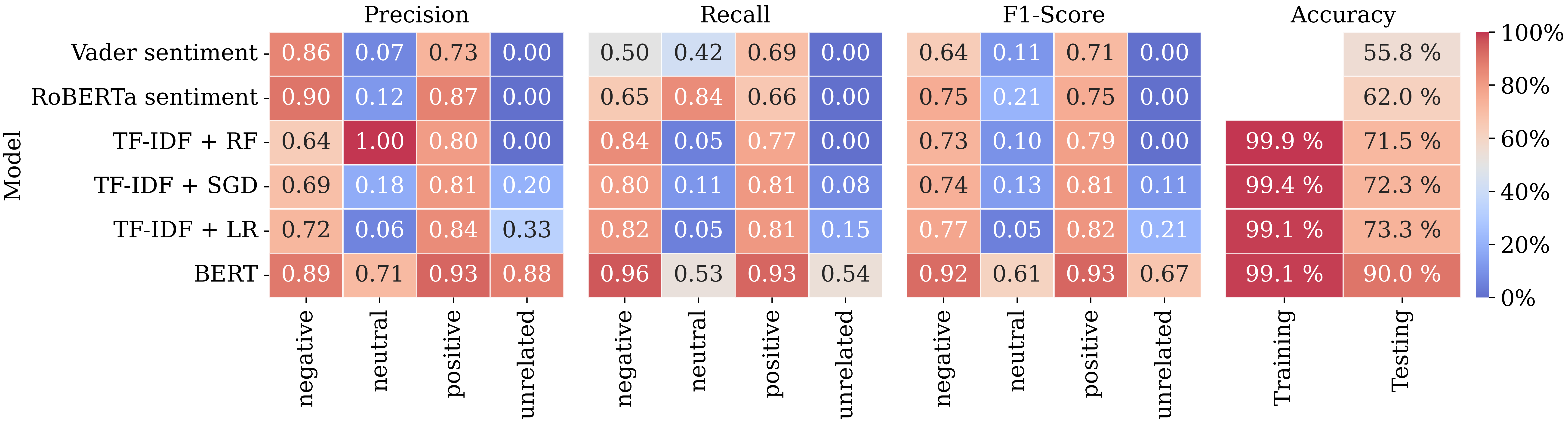}
  \caption{Classification performance of candidate models: Precision, Recall, F1-score, and Training and Testing accuracy.}
  \label{fig:performance}
\end{figure}

The six candidate models, after fine-tuning their hyperparameters, are evaluated based on four metrics: Precision, Recall, F1-score, and Accuracy. Accuracy represents the proportion of accurately classified cases out of all cases. Precision estimates the proportion of correctly identified positive cases out of all cases identified by the model. Recall estimates the proportion of correctly identified positive cases out of all relevant cases. F1-Score combines precision and recall. The model evaluation is conducted using the testing dataset. The performance of each model is presented in Figure \ref{fig:performance}.

\section{Lexical salience-valence analysis}
\label{secA3}

For textual analysis, we employ a technique called lexical salience-valence analysis (LSVA). Unlike simply counting the frequency of words in positive or negative reviews, LSVA enables to visualize the frequency of words across the entire corpus and assess their impact on overall sentiment \cite{taecharungroj2019analysing}. This method defines the salience and valence of a word as follows:

\begin{equation}
    \centering
    \text{salience}|_{\text{word}_i} = \log_{10}(N_{\text{total}})|_{\text{word}_i}
\end{equation}

\begin{equation}
    \centering
    \text{valence}|_{\text{word}_i} = \frac{N(\text{positive}) - N(\text{negative})}{N_{\text{total}}}|_{\text{word}_i}
\end{equation}

where
\begin{align*}
    N_{\text{total}} &\text{ represents the total number of reviews that } \text{word}_i \text{ appears} \\
    N_{\text{positive}} &\text{ denotes the number of positive reviews that } \text{word}_i \text{ appears} \\
    N_{\text{negative}} &\text{ denotes the number of negative reviews that } \text{word}_i \text{ appears}
\end{align*}

Salience is determined by taking the logarithm (base 10) of the frequency of each term (\(N_{\text{total}}\)). Valence is calculated by subtracting the number of negative occurrences (\(N_{\text{negative}}\)) from the number of positive occurrences (\(N_{\text{positive}}\)) and then dividing by the total count of the word (\(N_{\text{total}}\)). This ratio indicates how positively a word is perceived in the corpus. Reviews containing words with high positive valence are more likely to be positive, whereas those with high negative valence are more likely to be negative.

\section{LSVA analysis for POI types}
\label{secA4}

\begin{figure}[htb]
  \centering
  \includegraphics[width=1\textwidth]{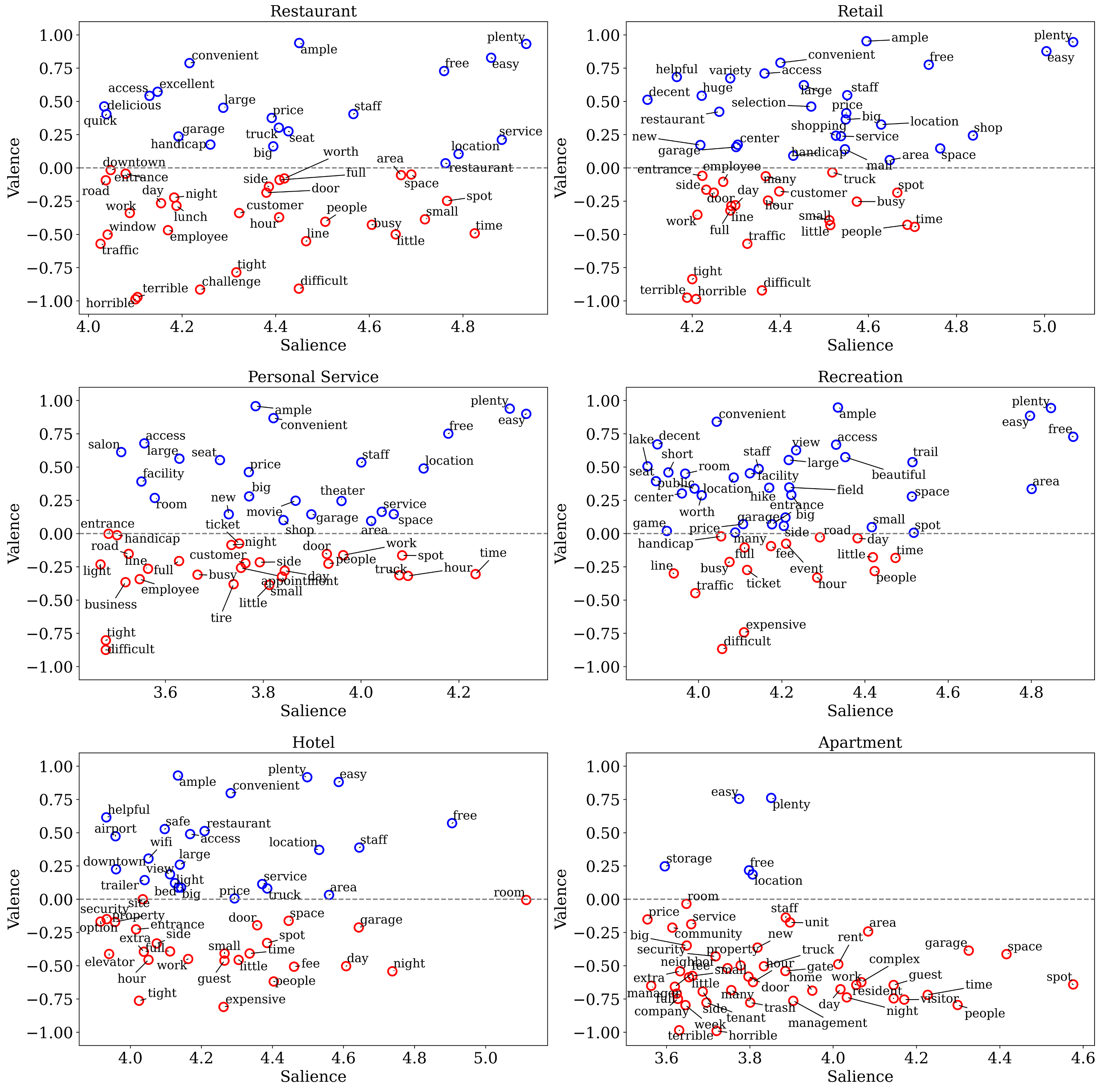}
  \caption{POI textual analysis}

  \label{fig:poi_textual}
\end{figure}

Figure \ref{fig:poi_textual} provides an overview of the terms associated with positive and negative sentiments in each of the six major POI types. This figure supports the analysis in Section 3.4 to understand the factors contributing to parking sentiment.

\section{Summary of socio-spatial variables}
\label{secA5}
Detailed information on all independent variables, along with their descriptive statistics (unstandardized), can be found in Table \ref{tab:variable_summary}. Variables in \textit{Italic} text are excluded from the models either due to their high multicollinearity or their limited capability to explain the dependent variables. The data sources used in our study are as follows: Socioeconomic and demographic factors are obtained from the 2021 5-year American Community Survey (ACS) \cite{ACS2022} conducted by the U.S. Census Bureau. Job-related variables are obtained from the 2020 Longitudinal Employer-Household Dynamics (LEHD) \cite{LEHD2020} data products. POI-related variables are obtained from Google Maps. Partisanship (county-level) is retrieved from the 2020 presidential election results provided by the MIT election lab. Other factors, such as road density, walkability, and transit frequency, are acquired from the Smart Location Database (SLD) published by the U.S. Environmental Protection Agency (EPA). 

\begin{table*}[]
\centering
\caption{Summary of CBG-level independent variables}
\label{tab:variable_summary}
\resizebox{\textwidth}{!}{%
    \begin{tabular}{llll}
    \hline
    \textbf{Variable} & \textbf{Description} & \textbf{Mean} & \textbf{Std.} \\ \hline
    \textbf{Socioeconomics} &  &  &  \\ \hline
    Population Density & Gross population density (people/acre) & 9.033 & 16.61 \\ \hline
    Employment Density & Gross employment density (jobs/acre) & 8.042 & 80.461 \\ \hline
    Poverty & \% households below poverty & 13.666 & 12.576 \\ \hline
    Rural Population & \% rural population & 10.198 & 26.475 \\ \hline
    \textit{Urban Population} & \% \textit{urbanized population} & 82.265 & 36.423 \\ \hline
    \textit{Median Income} & \textit{Median household income (Inflation-Adjusted), in \$ $10^{3}$/household} & 7.046 & 3.643 \\ \hline
    Highly-Educated & \% residents with education degree equal to or higher than Bachelor & 35.340 & 21.367 \\ \hline
    \textit{Democrat} & \% \textit{Democrats in 2020 presidential candidate vote totals} & 51.637 & 16.143 \\ \hline
    Zero Car & \% zero-car households & 9.646 & 12.921 \\ \hline
    \textit{One Car} & \% \textit{one-car households} & 36.144 & 15.222 \\ \hline
    \textit{\textgreater{}=2 Cars} & \% \textit{two-plus-car households} & 53.976 & 21.034 \\ \hline
    \textbf{Demographics} &  &  &  \\ \hline
    Male & \% males & 49.164 & 6.580 \\ \hline
    Age 18-44 & \% residents between 18 and 44 years & 37.883 & 14.222 \\ \hline
    Age 45-64 & \% residents between 45 and 64 years & 25.517 & 8.503 \\ \hline
    Age over 65 & \% residents 65 years and over & 16.404 & 10.859 \\ \hline
    \textit{White} & \textit{\% Non-Hispanic Whites }& 60.234 & 28.817 \\ \hline
    Asian & \% Asians & 6.343 & 10.883 \\ \hline
    African American & \% African Americans & 11.430 & 18.396 \\ \hline
    Hispanic & \% Hispanics/Latinos & 18.582 & 22.540 \\ \hline
    Others & \% Other minorities. & 0.746 & 3.451 \\ \hline
    \textbf{Land development} &  &  &  \\ \hline
    \textit{POI Density} & \textit{Total POI density (count/square miles) }& 43.857 & 90.965 \\ \hline
    Road Density & Total road density (mile/square miles) & 17.748 & 10.224 \\ \hline
    Parking POI Density & Parking lots and garage density (count/square miles) & 1.030 & 8.031 \\ \hline
    Walkability & \begin{tabular}{@{}l@{}}The combination of four measures: employment and household \\ entropy, static 8-tier employment entropy, street intersection \\ density, and distance to nearest transit stop. Ranked from 1 \\ (lowest support for walking) to 20 (highest support). \end{tabular}  & 11.649 & 4.185 \\ \hline
    Transit Frequency & Aggregate frequency of transit service per hour & 10.021 & 30.479 \\ \hline
    Avg. POI Score & Average POI score from Google Map Reviews & 4.279 & 0.232 \\ \hline
    \end{tabular}%
}
\end{table*}

\begin{figure}[h]
  \centering
  \includegraphics[width=0.7\textwidth]{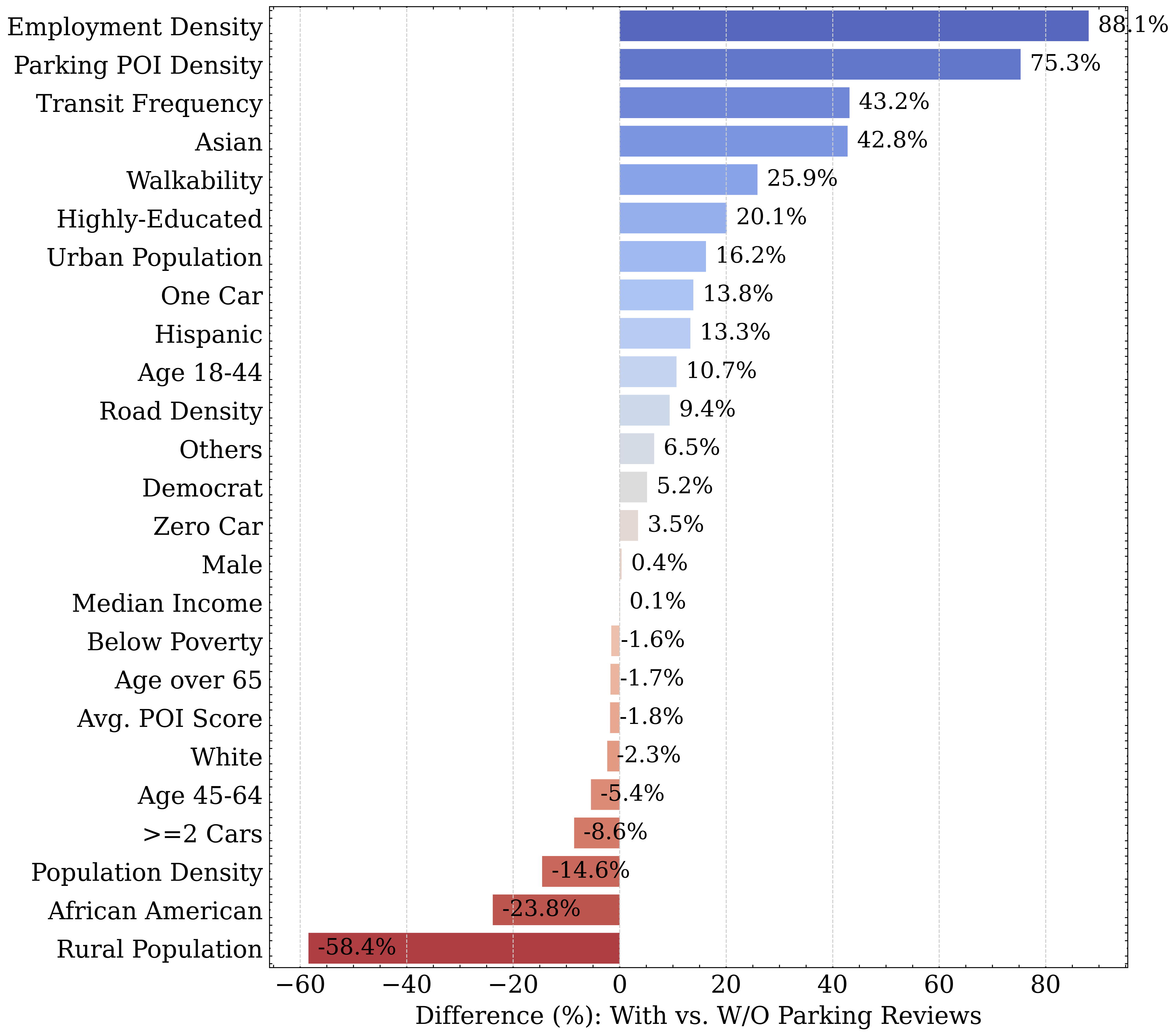}
  \caption{Socio-spatial disparities in parking review engagement: Comparing CBGs with and without parking reviews.}
  \label{fig:diff_socio}
\end{figure}

To understand potential sampling biases, we examine differences in socio-spatial factors between CBGs with and without parking-related reviews. As shown in Figure \ref{fig:diff_socio}, areas characterized by higher commercial and job densities, such as higher employment and POI densities, more frequent transit services, and lower percentage of rural populations, are more likely to generate parking reviews. Notable racial disparities also emerge, particularly among Asian and African American populations, which may reflect the location preferences associated with these groups' activities. Another interesting finding is the 14.6\% lower population density in areas with parking reviews compared to those without. This may be attributed to the absence of parking reviews for residential areas on Google Maps. Therefore, regions with high residential densities but lacking employment or service sectors tend not to produce reviews on Google Maps. In summary, this analysis highlights a major limitation of this study: parking sentiment is primarily linked to commercial and workplace activities, while those related to residential parking are not adequately captured.

\section{LISA cluster of parking sentiment}
\label{secA6}
\begin{figure}[h]
  \centering
  \includegraphics[width=0.7\textwidth]{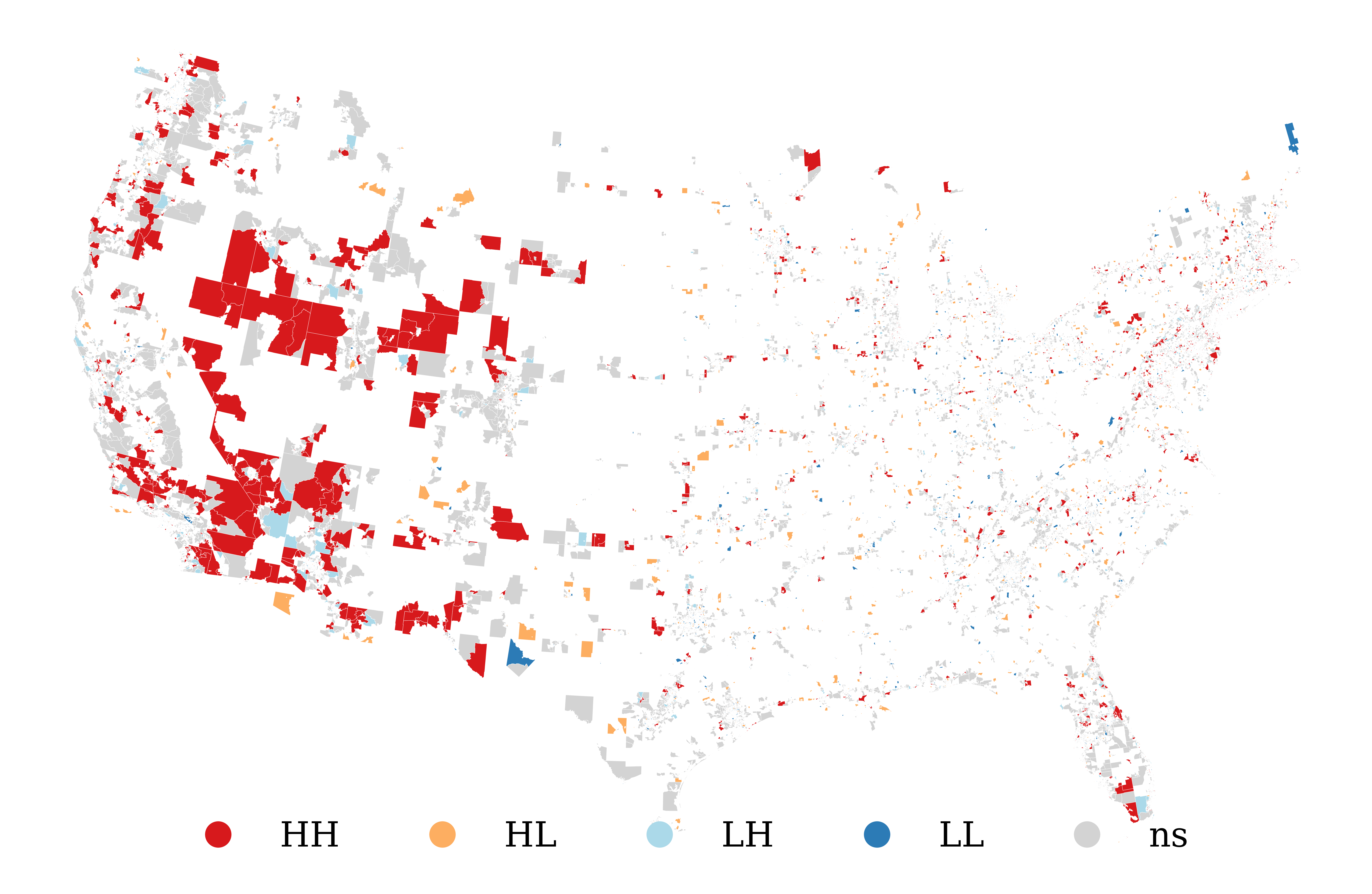}
  \caption{LISA cluster map of parking sentiment. Colors indicate significant positive (red and blue), negative (orange and pale blue) and not significant (gray) spatial autocorrelation.}
  \label{fig:spatial_cluster}
\end{figure}

We employ the Local Indicators of Spatial Association (LISA) cluster analysis \cite{anselin1995local} to identify spatial clusters of parking sentiment. Each location's LISA value is calculated based on its individual contribution to the global Moran's I statistic (0.173, P < 0.001). It provides a nuanced perspective to help identify areas with homogeneously high (HH), homogeneously low (LL), as well as heterogeneous (HL and LH) sentiment spatial patterns. As depicted in Figure \ref{fig:spatial_cluster}, most regions are marked in grey, indicating that parking sentiment in these areas does not exhibit significant spatial dependency. Nonetheless, a notable proportion of regions, particularly in the Rocky Mountains and Southwest, are highlighted in red, signifying areas with significant positive spatial correlation. Other clusters, including the LL (in blue), HL (in orange), and LH (in light blue), are less concentrated but dispersed across the US. Overall, although the overall spatial correlation of parking sentiment is not markedly high, it remains statistically significant and requires careful consideration in analyses.

\section{Generalized additive model (GAM)}
\label{secA7}

To handle both linear and nonlinear effects such as regional random effects and spatial autocorrelation, we employ the generalized additive model (GAM) in this study:

\begin{equation}
g(E(Y)) = \beta_0 + \sum_{k=1}^{N_l} \beta_k X_k + ti(\text{SR}) + s(\text{RE}) + \epsilon
\end{equation}

where \(Y\) is one of the seven aforementioned dependent variables, assuming to follow Gaussian distribution; \(g(\cdot)\) is the link function (identity function here); \(\beta_0\) is the overall intercept; \(\beta_k\) is the coefficient of the \(k\)th variable \(X_k\), and \(N_l\) is the number of variables with linear effects. Here \(X_k\) includes all socio-spatial factors listed in \ref{tab:variable_summary}; \(SR\) is the spatial autocorrelation, which is fitted by \(ti(\cdot)\), a marginal nonlinear smoother that excludes the basic functions associated with the main effects; \(RE\) is the CBSA-level random effects, which are captured via the penalized spline function \(s(\cdot)\); \(\epsilon\) is the error term, accounting for unexplained variation in the model. 

\begin{figure}[h]
  \centering
  \includegraphics[width=0.7\textwidth]{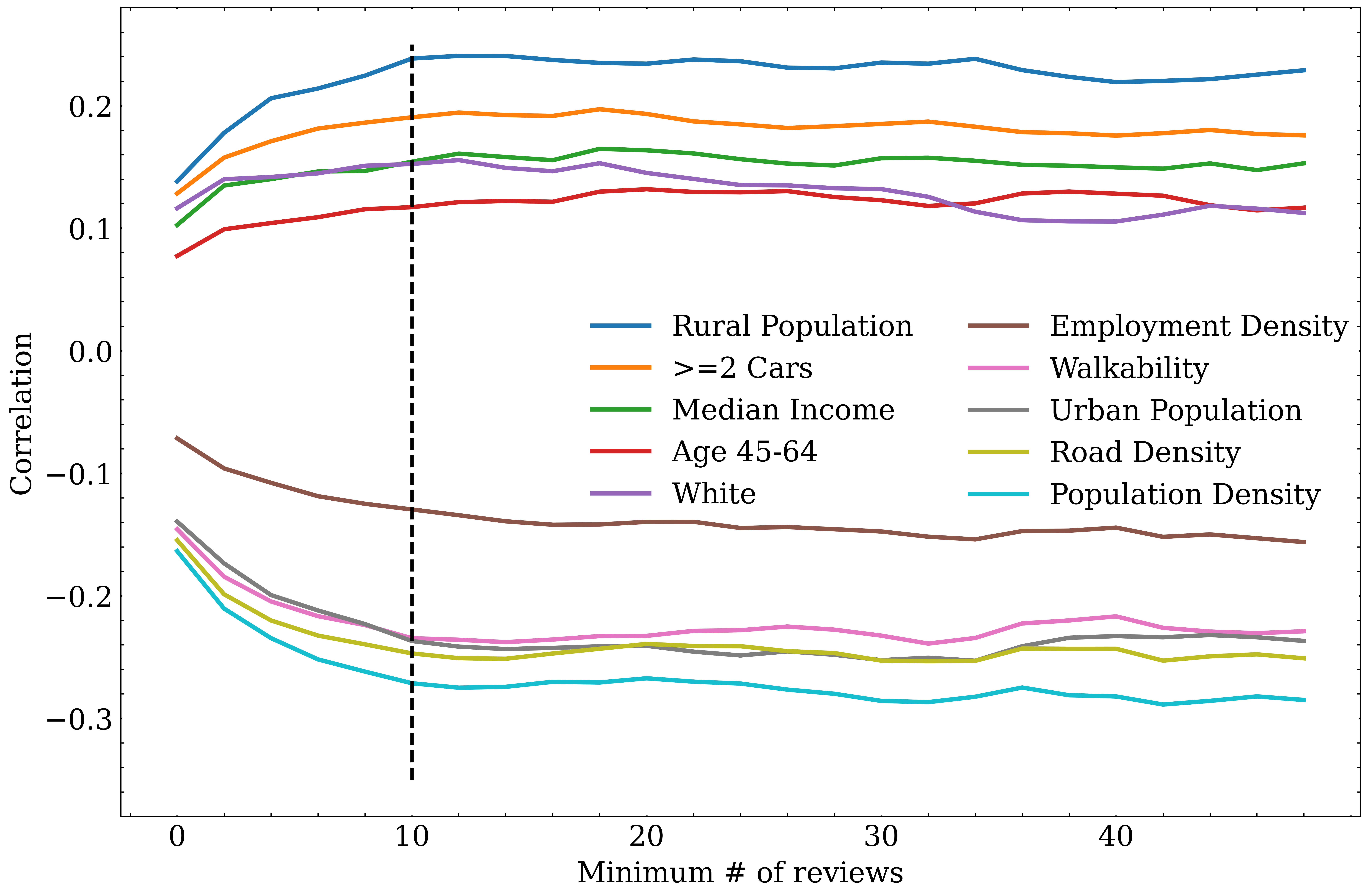}
  \caption{Sensitivity analysis on minimum number of reviews}
  \label{fig:min_thred}
\end{figure}

In this study, we use a minimum threshold of 10 parking reviews to select samples for model fitting. Alternatively, regions with fewer than 10 parking reviews are excluded from our regression analysis. Figure \ref{fig:min_thred} presents a sensitivity analysis that varies the minimum threshold from 0 to 50, exploring how changes in the threshold impact the correlation between parking sentiment and socio-spatial factors. As shown, across most factors, the correlations, whether positive or negative, are initially less pronounced and tend to stabilize as the minimum number of reviews increases. This pattern suggests that areas with fewer reviews may be less representative or more likely to be outliers, thereby diminishing the overall correlation. We finally choose 10 as the threshold, marked by the vertical dashed line, since this is the point where the correlations for most factors begin to stabilize. This provides an optimal balance between data inclusion and reliability.

\end{appendices}

\bibliography{sn-bibliography}

\end{document}